
\input phyzzx
\def\Landscapeout{\True}  


\font\elevenrm=cmr10 scaled\magstephalf
\def\sgamma{{\rm S$\Gamma$}}
\def\SPL{{\rm SPL(2,I$\!$R)}}
\def\cplain{{\rm C}\!\!\!{\rm\elevenrm l}}
\def\chx{\cosh X}

\def\eichi{\epsilon_1}
\def\eni{\epsilon_2}
\def\ebichi{\bar\epsilon_1}
\def\ebni{\bar\epsilon_2}
\def\I{({\rm I})}
\def\Ini{({\rm I}\!{\rm I})}
\def\Isan{({\rm I}\!{\rm I}\!{\rm I})}
\def\dela{{\partial}_A}
\def\delb{{\partial}_B}

\def\delzb{{\partial}_{\bar z}}

\def\delx{{\partial}_x}
\def\dely{{\partial}_y}
\def\lap{{\bigtriangleup}_{\rm SLB}}
\def\sqr#1#2{{\vcenter{\hrule height.#2pt
      \hbox{\vrule width.#2pt height#1pt \kern#1pt
          \vrule width.#2pt}
      \hrule height.#2pt}}}
\def\square{{\mathchoice{\sqr84}{\sqr84}{\sqr{5.0}3}{\sqr{3.5}3}}}
\def\hako{\square_{\,0}}

\def\signk{\sigma_k}
\def\edelv{e^{-\left({1-a\over{2a}}\right)^2\tau}}

\def\sgsum{\sum_{g\in{\rm S}\Gamma}}

\def\sumprim{\sum_{p\in{\rm Prim}({\rm S}\Gamma)}}
\def\proprim{\prod_{p\in{\rm Prim}({\rm S}\Gamma)}}
\def\lBn{\lambda^B_n}
\def\lFn{\lambda^F_n}

%
\catcode`\@=11 
\paperfootline={\hss\iffrontpage\else\ifp@genum\tenrm
    -- \folio\ --\hss\fi\fi}
\def\titlestyle#1{\par\begingroup \titleparagraphs
     \iftwelv@\fourteenpoint\fourteenbf\else\twelvepoint\twelvebf\fi
   \noindent #1\par\endgroup }
\def\eqnalign{\eqname}
\def\GENITEM#1;#2{\par \hangafter=0 \hangindent=#1
    \Textindent{#2}\ignorespaces}
\def\papersize{\hsize=35pc \vsize=52pc \hoffset=0.5pc \voffset=0.8pc
   \advance\hoffset by\HOFFSET \advance\voffset by\VOFFSET
   \pagebottomfiller=0pc
   \skip\footins=\bigskipamount \normalspace }
\papers  
\def\address#1{\par\kern 5pt \titlestyle{\twelvepoint\sl #1}}
\def\abstract{\par\dimen@=\prevdepth \hrule height\z@ \prevdepth=\dimen@
   \vskip\frontpageskip\centerline{%
	\iftwelv@\fourteencp\else\twelvecp\fi Abstract}\vskip\headskip }
\newif\ifYITP \YITPtrue
\font\fourteenmib =cmmib10 scaled\magstep2    \skewchar\fourteenmib='177
\font\elevenmib   =cmmib10 scaled\magstephalf   \skewchar\elevenmib='177
\def\YITPmark{\hbox{\fourteenmib YITP\hskip0.2cm
        \elevenmib Uji\hskip0.15cm Research\hskip0.15cm Center\hfill}}
\def\titlepage{\FRONTPAGE\papers\ifPhysRev\PH@SR@V\fi
    \ifYITP\null\vskip-1.70cm\YITPmark\vskip0.6cm\fi 
   \ifp@bblock\p@bblock \else\hrule height\z@ \rel@x \fi }
\catcode`\@=12 
\newbox\leftpage \newdimen\fullhsize
\catcode`\@=11 
\ifx\Landscapeout\Undefined\message{(This is unreduced.}
\else
\tolerance=1000\hfuzz=2pt
\message{(This is the preprint format.} \let\l@r=L
\def\refout{\par\penalty-400\vskip\chapterskip
   \spacecheck\referenceminspace
   \ifreferenceopen \Closeout\referencewrite \referenceopenfalse \fi
   \line{\twelverm\hfil REFERENCES\hfil}\vskip\headskip
   \input \jobname.refs
   }
\def\figout{\par\penalty-400
   \vskip\chapterskip\spacecheck\referenceminspace
   \iffigureopen \Closeout\figurewrite \figureopenfalse \fi
   \line{\twelverm\hfil FIGURE CAPTIONS\hfil}\vskip\headskip
   \input \jobname.figs
   }
\magnification=1000\baselineskip=15pt plus 0.2pt minus 0.1pt\vsize=7truein
\fullhsize=11truein
\def\papersize{\hsize=4.9truein\vsize=7truein\hoffset=-0.3in\voffset=-0.5in
	\pagebottomfiller=0pc \skip\footins=\bigskipamount  }
\paperfootline={\hss\iffrontpage\else\ifp@genum\ninerm
	-- \the\count0\ -- \hss\fi\fi}
\font\twelvemib   =cmmib10 scaled\magstep1	    \skewchar\twelvemib='177
\font\tenmib      =cmmib10			    \skewchar\tenmib='177
\def\YITPmark{\hbox{\twelvemib YITP\hskip0.2cm
        \tenmib Uji\hskip0.15cm Research\hskip0.15cm Center\hfill}}
\def\titlepage{\FRONTPAGE\papers\ifPhysRev\PH@SR@V\fi
    \ifYITP\null\vskip-1cm\YITPmark\vskip0.6cm\fi 
   \ifp@bblock\p@bblock \else\hrule height\z@ \rel@x \fi }
\output={\almostshipout{\leftline{\vbox{\pagebody\makefootline}}}
	\advancepageno}
\def\almostshipout#1{\if L\l@r \count1=1 \message{[\the\count0.\the\count1]}
      \global\setbox\leftpage=#1 \global\let\l@r=R
 \else \count1=2
  \shipout\vbox{{\hsize\fullhsize\makeheadline}
      \hbox to\fullhsize{\box\leftpage\hfil#1\hss}}  \global\let\l@r=L\fi}
\tenpoint
\fi
\catcode`\@=12 
%
\catcode`\@=11 
\let\rel@x=\relax
\let\n@expand=\relax
\def\pr@tect{\let\n@expand=\noexpand}
\let\protect=\pr@tect
\let\gl@bal=\global
%
%
\newtoks\t@a \newtoks\t@b \newtoks\next@a
\newcount\num@i \newcount\num@j \newcount\num@k
\newcount\num@l \newcount\num@m \newcount\num@n
\long\def\l@append#1\to#2{\t@a={\\{#1}}\t@b=\expandafter{#2}%
                         \edef#2{\the\t@a\the\t@b}}
\long\def\r@append#1\to#2{\t@a={\\{#1}}\t@b=\expandafter{#2}%
                         \edef#2{\the\t@b\the\t@a}}
\def\l@op#1\to#2{\expandafter\l@opoff#1\l@opoff#1#2}
\long\def\l@opoff\\#1#2\l@opoff#3#4{\def#4{#1}\def#3{#2}}
\newif\ifnum@loop \newif\ifnum@first \newif\ifnum@last
\def\sort@@#1{\num@firsttrue\num@lasttrue\sort@t#1}
\def\sort@t#1{\pop@@#1\to\num@i\rel@x
            \ifnum\num@i=0 \num@lastfalse\let\next@a\rel@x%
            \else\num@looptrue%
                 \loop\pop@@#1\to\num@j\rel@x
                    \ifnum\num@j=0 \num@loopfalse%
                    \else\ifnum\num@i>\num@j%
                         \num@k=\num@j\num@j=\num@i\num@i=\num@k%
                         \fi%
                    \fi%
                    \push@\num@j\to#1%
                  \ifnum@loop\repeat%
                  \let\next@a\sort@t%
            \fi%
            \print@num%
            \next@a#1}
\def\print@num{%
              \ifnum@first%
                 \num@firstfalse\num@n=\num@i\number\num@i%
              \else%
                 \num@m=\num@i\advance\num@m by-\num@l%
                 \ifcase\num@m\message{%
                   *** WARNING *** Reference number %
                   [\the\num@i] appears twice or more!}%
                 \or\rel@x%
                 \else\num@m=\num@l\advance\num@m by-\num@n%
                    \ifcase\num@m\rel@x%
                    \or,\number\num@l%
                    \else-\number\num@l%
                    \fi%
                    \ifnum@last\num@n=\num@i,\number\num@i\fi%
                 \fi%
              \fi%
              \num@l=\num@i%
              }
\def\pop@@#1\to#2{\l@op#1\to\z@@#2=\z@@}
\def\push@#1\to#2{\edef\z@@{\the#1}\expandafter\r@append\z@@\to#2}
\def\append@cs#1=#2#3{\xdef#1{\csname%
                    \expandafter\g@bble\string#2#3\endcsname}}
\def\g@bble#1{}
\def\if@first@use#1{\expandafter\ifx\csname\expandafter%
                              \g@bble\string#1text\endcsname\relax}
\def\keep@ref#1#2{\def#1{0}\append@cs\y@@=#1{text}\expandafter\def\y@@{#2}}
\def\keepref#1#2{\if@first@use#1\keep@ref#1{#2}%
                 \else\message{%
                    \string#1 is redefined by \string\keepref! %
                    The result will be .... what can I say!!}%
                 \fi}
\def\Null{0}
\def\get@ref#1#2{\def#2{\string#1text}}
\def\findref@f#1{%
                \ifx#1\Null \get@ref#1\text@cc\R@F#1{{\text@cc}}%
                \else\rel@x\fi}
\def\findref#1{\findref@f#1\ref@mark{#1}}
\def\find@rs#1{\ifx#1\endrefs \let\next=\rel@x%
              \else\findref@f#1\r@append#1\to\void@@%
                    \let\next=\find@rs \fi \next}
\def\findrefs#1\endrefs{\def\void@@{}%
                    \find@rs#1\endrefs\r@append{0}\to\void@@%
                    \ref@mark{{\sort@@\void@@}}}
\let\endrefs=\rel@x
\def\ref@mark#1#2{\if#2,\rlap#2\refmark#1%
                  \else\if#2.\rlap#2\refmark#1%
                   \else\refmark{{#1}}#2\fi\fi}
\def\NPref@mark#1#2{\if#2,\NPrefmark#1,%
                  \else\if#2.\NPrefmark#1.%
                   \else\thinspace\NPrefmark{{#1}} #2\fi\fi}
\def\R@F#1{\REFNUM #1\R@F@WRITE}
\def\NPrefitem#1{\r@fitem{[#1]}}
\def\NPrefs{\let\refmark=\NPrefmark \let\refitem=\NPrefitem%
            \let\ref@mark=\NPref@mark}
\def\PRrefs{\let\refmark=\PRrefmark}
\def\R@F@WRITE#1{\ifreferenceopen\else\gl@bal\referenceopentrue%
     \immediate\openout\referencewrite=\jobname.refs%
     \toks@={\begingroup \refoutspecials}%
     \immediate\write\referencewrite{\the\toks@}\fi%
    \immediate\write\referencewrite{\noexpand\refitem%
                                    {\the\referencecount}}%
    \immediate\write\referencewrite#1}
\catcode`\@=12 

\Pubnum={YITP/U-93-14}
\date={May 1993}

\titlepage
\title{Chaotic System on the Super Riemann Surface}

\author{Shuji Matsumoto
\foot{E-mail address: shuji@kekvax.kek.jp}}
\address{KEK, National Laboratory for High Energy Physics, Oho, Tsukuba,
Ibaraki 305, Japan}
\author{Shozo Uehara
\foot{E-mail address: uehara@yisun1.yukawa.kyoto-u.ac.jp}}
\address{Uji Research Center, Yukawa Institute for Theoretical Physics,
Kyoto University, Uji 611, Japan}
\andauthor{Yukinori Yasui
\foot{E-mail address: f51999@sakura.kudpc.kyoto-u.ac.jp}}
\address{Department of Physics, Osaka City University, Sumiyoshi, Osaka
558, Japan}

\abstract{
A chaotic system of a nonrelativistic superparticle moving freely on
the super Riemann surface (SRS) of genus $g \geq 2$ is reviewed.
}
\endpage

\sequentialequations
\NPrefs
\keepref\hadamard{ J~.Hadamard, J. Math. Pure Appl. {\bf 4} (1898) 27.}
\keepref\balazs{ for a recent review, see N.L.~Balazs, and A.~Voros,
	Phys. Rep. {\bf 143} (1986) 109.}
\keepref\selberg{A.~Selberg, J. Indian Math. Soc. {\bf 20} (1956) 47.}
\keepref\hejhal{D.A.~Hejhal, {\sl Lecture Note in Mathematics},
	Vol.{\bf 548}; Berlin,Heidelberg, New~York: Springer 1976.}
\keepref\dhoker{E.~D'Hoker and D.H~Phong, Nucl. Phys. {\bf B269}
	(1986) 205; Phys. Rev. Lett. {\bf 56} (1986) 912.}
\keepref\berezin{F.A.~Berezin,{\sl Introduction to Superanalysis},
	Math. Phys. and Appl. Math. Vol.{\bf 9}, : D.~Reidel
	Publishing 1987.}
\keepref\dewitt{B.~DeWitt, {\sl Supermanifold}, Cambridge Univ. Press 1984}
\keepref\friedan{D.~Friedan, {\sl in the Proceedings of the Workshop
	on Unified Theories}, edited by D.~Gross and M.~Green,
	Shingapore: World Scientific 1986.}
\keepref\ourfirst{S.~Matsumoto and Y.~Yasui, Prog. Theor. Phys.
	{\bf 79} (1988) 1022.}
\keepref\oursecond{S.~Uehara and Y.~Yasui,Phys. Lett. {\bf 202B}
	(1988) 530.}
\keepref\ourthird{S.~Uehara and Y.~Yasui, J. Math. Phys. {\bf 29}
	(1988) 2486.}
\keepref\ourfourth{S.~Matsumoto, S.~Uehara and Y.~Yasui, Phys. Lett.
	{\bf 134A} (1988)  81.}
\keepref\ourfifth{S.~Matsumoto, S.~Uehara and Y.~Yasui, J. Math. Phys.
	{\bf 31} (1990) 476.}
\keepref\crane{L.~Crane and J.M.~Rabin, Comm. Math. Phys. {\bf 100}
	(1985) 141; {\bf 113} (1988) 601.}
\keepref\aoki{K.~Aoki, Comm. Math. Phys. {\bf 117} (1988) 405.}
\keepref\howe{P.~Howe, J. Phys. {\bf A12} (1979) 393.}
\keepref\baranov{A.M.~Baranov,Yu.I.~Manin, I.V.~Frolov and
	A.S.~Schwarz, Comm. Math. Phys. {\bf 111} (1987) 373.}
\keepref\anosov{D.V.~Anosov, Proc. Steklov Inst. of Math. {\bf 90} (1967).}
\keepref\sinai{Ya.G.~Sinai, {\sl Introduction to Ergodic Theory},
	Princeton University Press 1976.}
\keepref\pesin{Ya.V.~Pesin, Dolk. Akad. Nauk SSSR {\bf 226} (1976)
	(Soviet Math. Dolk. {\bf 17} (1976) 196).}
\keepref\omote{M.~Omote and H.~Sato, Prog. Theor. Phys. {\bf 47}
	(1972) 1367.}
\keepref\kawai{T.~Kawai, Foundation of Phys. {\bf 5} (1975) 143.}
\keepref\oursixth{S.~Uehara and Y.~Yasui, Phys. Lett. {\bf 217B}
	(1989)  479; Comm. Math. Phys. {\bf 144} (1992) 53.}
\keepref\grosche{C.~Grosche, Comm. Math. Phys. {\bf 133} (1990)
	433; {\bf 151} (1993) 1.}
\keepref\sunada{T.~Sunada, {\sl Kihon-gun to Laplacian}
	Kinokuniya Suugaku Sensho 29; Kinokuniya (1988).}

\chapter{Introduction}

 In the present paper, we review a supersymmetric extension of the
Hadamard model, the classical and quantum motions of a superparticle on
the super Riemann surface (SRS).

 The conventional Hadamard model\findref\hadamard represents the free
motion of a nonrelativistic particle on the compact Riemann surface of
a constant negative curvature\findref\balazs.
The classical motion is known for its strongly
chaotic property.  The Hadamard model was originally proposed as a simple
example of the dynamical system which possesses the ergodic property.
This model has various advantages in the study of the chaotic
properties mathematically.  One of the essential features of this
model is that there exists a well-defined
quantum dynamics where the Laplace-Beltrami operator on the Riemann
surface acts as the Hamiltonian.
	The model may be useful in physics to examine the properties
of a quantum chaos.
	Specifically, the model shows us the relation between a
classical and a quantum chaos.
	A quantized energy sum rule is actually the celebrated Selberg
trace formula\findrefs\selberg\hejhal\endrefs.
	The quantum energy spectrum is complicated, however, it would
be obtained through the Selberg zeta function.

	Here, we intend to bring the supersymmetry into the Hadamard
model. That is, we investigate the system of a superparticle moving
freely on a compact super Riemann surface of genus $g\geq 2$.
The supersymmetrical Hadamard model will offer
\pointbegin
an application of the superanalog of the analytic theories
on a Grassmann algebra\findref\berezin.
\point
an example of integrable classical and quantum dynamical systems with
supersymmetry (the motion on SH, a universal covering space of the
SRS, is expected to be integrable).
\point
the notion of supersymmetrized chaos (however it seems to be rather
puzzling).
\point
superanalogs of the trace formula and the zeta function, which are
important for the superstring theory (the notion of an SRS comes
naturally in the superspace approach of superstrings\findref\friedan).

	Motivated by them and armed with the mathematical tools for the
supersymmetry, we develop the theory along the conventional study of the
Hadamard model.
	This paper is organized as follows.
In the next section the notations and the conventions of super Riemann
surfaces are presented and the Lagrangian of a superparticle on a
super Riemann surface is given.
	Section 3 is devoted to the classical mechanics for the system
and quantization is carried out in Sec. 4.
	Superanalogs of the Selberg trace formula and the zeta
function are given in Sec. 5.  Section 6 is devoted to the discussions
on the classical chaos.  In the final section we comment on the moduli
space of the super Riemann surface.
	This paper is based on our previous
works\findrefs\ourfirst\oursecond\ourthird\ourfourth\ourfifth\endrefs.

	\chapter{Preliminaries}

	This section is devoted to compiling the notations and
conventions of super Riemann surfaces (SRSs) of genus $g\geq2$.
As in the conventional Hadamard model, we will  employ here the
convenient way to represent the SRS.  We will see that it is
represented as a fundamental domain of a certain universal covering
space.  And this brings us an advantage that we can investigate the
motions on an SRS by imposing a periodic boundary condition on the
motions on the covering space.  The free motion on an SRS is supposed
to be generated by the {\it distance-proportional} Lagrangian.  We
will give a metric tensor on the covering space parametrized by a real
parameter $a$ and also give the Lagrangian for a superparticle based
on the distance.

	As is well known, the Poincar\'e upper half-plane H $=\{z\in
\cplain\ | {\rm Im}\,z >0\}$ together with the group
of its conformal automorphisms PSL(2,I$\!$R) (M\"obius group) is a
model of the hyperbolic geometry.  Turning to the supergeometry, we
can extend all the standard constructions to the superplanes.  The
super Poincar{\'e} upper half-plane SH is expressed by one Grassmann
even and one odd complex coordinate $z$ and $\theta$,respectively;
\foot{Im$\,z>0$ means that Im$\,z_0>0$ with $z_0$ being the body part
of $z$. We shall use such a convention for inequalities through out
this paper for simplicity. }
$$
	{\rm SH} = \{ Z = (z,\theta)\ |\ {\rm Im}\,z>0\}.     \eqn\ichi
$$
The superconformal automorphisms \SPL\  of SH consist of such
transformation $k:(z,\theta)\mapsto(\tilde z,\tilde\theta)$ as
$$
\eqalign{
\tilde z&={{az+b}\over{cz+d}}+\theta{{\alpha z
	+\beta}\over{(cz+d)^2}}, \cr \tilde\theta&= {{\alpha
	z+\beta}\over{cz+d}}+\theta
	{{1+{1\over2}\beta\alpha}\over{cz+d}},\cr}
	\eqn\ni
$$
where $a,b,c\ {\rm and}\ d$ are Grassmann even and $\alpha$
and $\beta$ are Grassmann odd parameters with
\foot{As to complex conjugation, we adopt such a
	convention that $\bar\alpha = i\alpha, \bar\beta = i\beta.$}
$$
	ad-bc=1 ,~~~~~~~~~~ a,b,c,d \in{\rm I}\!{\rm R}. \eqn\san
$$
The above transformation \ni\ is, of course, superanalytic and is
also superconformal,
$$
\eqalignno{{}& D\tilde z - \tilde\theta D\tilde\theta =0~,
	&\eqnalign{\yon}\cr
	{}& D \equiv {\partial\over{\partial\theta}}
	+ \theta{\partial\over{\partial z}}~.&\eqnalign{\go}\cr}
$$
If we introduce homogeneous coordinates ($z_1,z_2,\xi$) of the complex
projective space, we can rewrite \ni~ as a linear transformation
($z=z_1 z_2^{-1}, \theta = \xi z_2^{-1}$),
$$
\eqalign{
      {}&\pmatrix{\tilde z_1\cr \tilde z_2\cr \tilde\xi\,\cr}
	= A_k \pmatrix{z_1\cr z_2\cr \xi\,\cr},\cr
	{}&    A_k = {\Big(}1+{1\over2}\beta\alpha{\Big)}^{-1} \times
   \pmatrix{ a      & b     & b\alpha-a\beta         \cr
             c      & d     & d\alpha-c\beta         \cr
             \alpha & \beta & 1+{3\over2}\beta\alpha \cr}
     ,~~~ {\rm sdet}\,A_k = 1.\cr }        \eqn\roku
$$
Similarly to the ordinary Riemann surfaces, every compact super
Riemann surface with genus $g\geq2$ can be represented as a quotient
space SH/\sgamma\findref\crane, where the universal covering space of
a super Riemann surface (SRS) is the super Poincar\'e upper half-plane
SH and the covering group \sgamma\ (called the super Fuchsian group)
is a discrete subgroup of superconformal automorphisms \SPL  having no
fixed points on SH.

	The super Fuchsian group S$\Gamma$ is generated by $2g$
elements $\{ A_i, B_i, i=1,\cdots,g\}$ satisfying a condition,
$$
     \prod_{i=1}^g \left( A_i B_i A_i^{-1} B_i^{-1}\right) = 1.
     \eqn\nana
$$
Each element of the generators contains three Grassmann even and two
odd parameters and the condition \nana\ is invariant under
$A_i\mapsto MA_iM^{-1},~~ B_i\mapsto  MB_iM^{-1}$
where $M\in$ \SPL.  Thus the set of the generators and hence the SRS
with genus $g\geq2$ are specified by $6g-6$ Grassmann even and $4g-4$
odd parameters. S$\Gamma$ acts discontinuously on SH and all its
elements are hyperbolic, i.e., the reduced subgroup, where odd
parameters are put to be zero, consists of the hyperbolic elements,
$|a+d|>2$.
Let Conj(\sgamma) be the set of all conjugacy classes of \sgamma\ and
${\rm Prim}({\rm S}\Gamma)=\{p\in{\rm Conj}({\rm S}\Gamma); p\ne
k^m\ {\rm for~any}~k\in{\rm Conj}({\rm S}\Gamma)\ {\rm and\ }m\ge2\}$
the set of all primitive conjugacy classes of \sgamma.  Then we have
$$
	Q({\rm Prim}({\rm S}\Gamma))={\rm Conj}({\rm S}\Gamma),~~{\rm
	where}~Q(P)=\{p^m,~p\in P,m\ge0\}. \eqn\kyu
$$

	An element $k\ne1$ of S$\Gamma$ causes such a transformation
as \ni. \sgamma\ acts effectively on SH, however, $k\in\!
{\rm S}\Gamma$ has two fixed points, $(u,\mu)$ and $(v,\nu)$,  on the
\lq\lq super'' real axis I$\!$R$_s\equiv
\{Z=(z,\theta)\ |\ {\rm Im}\,z=0,\ \bar\theta =i\theta\}$,
$$
	u,v={a-d\pm\sqrt{(a+d)^2-4}\over2c},~~~~\mu={{\alpha u+\beta}
	\over{cu+d-1}},~~ \nu={{\alpha v+\beta} \over{cv+d-1}}~.
	\eqn\juu
$$
These fixed points are repelling and attractive points, respectively,
$$
	\eqalign{k^{-n}&:~(z,\theta)\longrightarrow (u,\mu)~,\cr
	k^n&:~(z,\theta)\longrightarrow (v,\nu)~,
	\qquad{\rm for}~n\rightarrow\infty~.\cr}\eqn\ichiichi
$$
Let us define the quantities $N_k$ (norm function) and $\chi(k)$
(sign factor) for $k\in{\rm S}\Gamma$;
$$
	\chi(k)(N^{1/2}_k+N^{-1/2}_k)=a+d-{a+d+2\over2}\beta\alpha
	={\rm str}\,A_k+1.  \eqn\ichini
$$
Actually, $N_k(>1)$ is the square of the maximum eigenvalue of the
matrix $A_k$ and $\chi(k)$ has to be chosen as
$$
	\chi(k)=\cases{1~,&if~~~${\rm str}\,A_k+1>2$~;\cr -1~,&if~~~${\rm
	str}\,A_k+1<-2$~,\cr}
	\eqn\ichisan
$$
Using the transformation of \SPL, we see that any element $k\neq 1$ of
S$\Gamma$ is always conjugate in \SPL\ to the magnification
$$
  \eqalign{{\widetilde w}&=N_kw~,\cr
  {\tilde\eta}&=\chi(k)~N_k^{1/2}\eta~,
    \qquad N_k>1~,\cr} \eqn\ichigo
$$
or in the matrix representation:
$$
  A_f\,A_k\,A^{-1}_f=A_{fkf^{-1}}=
      {\rm diag}\left(\chi(k)N_k^{1/2},\chi(k)N_k^{-1/2},1\right)
	\equiv A_{mag}~.  \eqn\ichiroku
$$
Apparently, the magnification depends on the element of
${\rm Conj}({\rm S}\Gamma)$,
$$
    N_{gkg^{-1}}=N_k~,~~~~~\chi(g\,k\,g^{-1})=\chi(k)~.\eqn\ichinana
$$

	Now we will introduce a \SPL-invariant metric tensor
on SH which is a superanalog of the Poincar\'e metric on H.
The latter, $ds_0^2 = |dz|^2/({\rm Im}\,z)^2$, is invariant under
PSL(2,I$\!$R) and gives a constant negative curvature.
The corresponding volume element is
$$
   {dxdy\over{y^2}},\qquad\qquad ({\rm Re}\,z=x,\ {\rm Im}\,z=y).
 \eqn\ichikyu
$$
To construct the \SPL-invariant metric tensor on SH, we use the
techniques developed in the supergravity theory on $2+2$ dimensional
curved superspace.  The basic quantities are the super vielbein
$E_M^A$ which, however, are not completely independent
superfields.   It was shown that $2+2$ dimensional superspace is
superconformally flat\findref\howe\ where the basis one-forms
$\hat E^A$ are
$$
\hat E^{++}= dz+\theta d\theta,~~
\hat E^{--}=d\bar z-\bar\theta d\bar\theta,~~
\hat E^{+}= d\theta, ~~\hat E^-=d\bar\theta. \eqn\nijuu
$$
By the super Weyl transformation\findref\howe\ the vielbein
$\hat E^A_M$ changes as
$$
\eqalign{
\hat{E}^A_M\mapsto E^A_M=&
	\cases{
	E^a_M=\Lambda(Z)\hat{E}^a_M~,\cr
	E^\alpha_M=\Lambda^{1/2}(Z)\hat{E}^{\alpha}_M-i
	\hat{E}^a_M(\gamma_a)^{\alpha\beta}\hat
	{E}^N_{\beta} \partial_N\Lambda^{1/2}(Z),}\cr
{}&\hskip4cm(a=++,--, \alpha=+,-)\cr }
	\eqn\niichi
$$
where $\hat E^M_A=(\hat E^A_M)^{-1}$, $\Lambda(Z)$ is the scaling function and
$(\gamma_a)$ is the gamma matrix: $$
	(\gamma_+)^{\alpha\beta}=\pmatrix{0 & 2\cr 0 & 0},~~
	(\gamma_-)^{\alpha\beta}=\pmatrix{0 & 0\cr 2 & 0}. \eqn\nini
$$
We take a \SPL-covariant function for $\Lambda(Z)$,
$$
	\Lambda(Z)=Y^{-1}, \eqn\nisan
$$
where $Y$ is given by
$$
	Y={\rm Im}~z+{{1}\over{2}}\theta\bar\theta~,\eqn\niyon
$$
which is the superanalog of $y={\rm Im}\,z$ on H.
The $E^A$ are now given as
$$
  \eqalign{
  E^{++} &=Y^{-1}\,(dz+\theta d\theta)~,~E^{--} = \overline{E^{++}}~, \cr
  E^+\, &= Y^{-3/2}\left[
  (Y+{1\over2}\theta\bar\theta)d\theta+{1\over2}
  (i\theta-\bar\theta)dz  \right]~,~E^-\, = \overline{E^+}~. \cr} \eqn\ninana
$$
The \SPL-invariant line element can be constructed by
$$
	ds^2=E^{++}E^{--}-2a\,E^+E^-~, \eqn\nihachi
$$
where $a (\neq 0)$ is an arbitrary real Grassmann even number.  Rewriting
$ds^2=dq^A\,g_{AB}\,dq^B,
(q^z,q^{\bar z},q^{\theta},q^{\bar\theta}) =
(z,\bar z,\theta,\bar\theta)$,
we obtain the metric tensor on SH,
$$
\left(g_{AB}\right)=\pmatrix{
    0   &  {1\over2Y^2}  &  0   &
        -{\bar{\theta}+a(i\theta-\bar{\theta})\over 2Y^2}\cr
  {1\over 2Y^2} & 0  &  {\theta-a(\theta+i\bar{\theta})\over 2Y^2}
            &   0    \cr
  0  &   {a(\theta+i\bar{\theta})-\theta\over2Y^2}
   & 0 & {\theta\bar{\theta}-2a(Y+\theta\bar{\theta})\over 2Y^2}\cr
  {\bar{\theta}+a(i\theta-\bar{\theta})\over2Y^2}  &   0   &
   {2a(Y+\theta\bar{\theta})-\theta\bar{\theta}\over2Y^2}  &  0
   \cr}, \eqn\nikyu
$$
and the corresponding volume element is given by
$$
	dV={1\over{2aY}}\,dxdyd\theta d\bar\theta. \eqn\sanjuu
$$

	Since we have now a SPL(2,I$\!$R)-invariant line
element\nihachi, we give our Lagrangian of a superparticle with mass
$m$ on SH\findref\ourthird,
$$
   L={m\over 2}\left(ds\over dt\right)^2={m\over2}
	\dot q^A\,g_{AB}\,\dot q^B. \eqn\sanichi
$$
This is SPL(2,I$\!$R)-invariant and hence, of course,
S$\Gamma$-invariant, thus it is also the Lagrangian for a
superparticle on the SRS.

	\chapter{Classical Mechanics}

	In this section we examine the classical dynamics of a
superparticle on the SRS.
Firstly, we solve the motion on SH.  The motion on SH is
expected to be integrable.  According to the canonical theory of
supermechanics, if we find out the adequate number of integrals of
motion which {\it commute} each other with respect to the Poisson
bracket, we can construct the general solution out of the integrals.
However it is rather difficult to do it explicitly.  There is no
systematical way to get such integrals, and hence we take another
path.  We solve the Hamilton-Jacobi equation.
	The calculation is considerably cumbersome but the general
solution for the metric tensor with an arbitrary parameter $a$ is given.
Since the SRS is represented by the fundamental domain of SH,
SH/\sgamma, the motion on the SRS is given by imposing the
{\em periodic} boundary condition on the motion on SH. As we have
expected, the motion on the SRS shows chaotic properties.

	The Euler-Lagrange equations from $L$ in \sanichi~  are
geodesic equations,
$$
	\ddot q^A + \Gamma^A_{~BC}\,\dot q^C \dot q^B = 0,  \eqn\sanni
$$
where $\Gamma^A_{~BC}$ is the Cristoffel's symbol\findref\ourfifth.
Eq.\sanni~ is given explicitly,
$$
  \eqalign{{}& \ddot z+{1\over Y}(i\dot z^2-\dot z\dot{\theta}
  \bar{\theta})+{1-a\over2a}\left({i\over{Y^2}}\theta\bar{\theta}
  \dot z\dot{\bar z}-{2\over Y}\dot z\theta\dot{\bar{\theta}}\right)
  =0,\cr
  {}& \ddot{\theta}+{i\over Y}\dot z\dot{\theta}+{1-a\over 2a}
  \left({2Y+\theta\bar{\theta}\over{Y^2}}\dot z\dot{\bar{\theta}}
  -{\theta+i\bar{\theta}\over{Y^2}}\dot z\dot{\bar z}
  +{2\over Y}\theta\dot{\theta}\dot{\bar{\theta}}
  +{i\over{Y^2}}\dot{\bar z}\theta\bar{\theta}\dot{\theta}\right)
  =0, \cr} \eqn\sansan
$$
and their complex conjugated ones. The body part of the
Eqs.\sansan\ is
$$
	\ddot z_0 + {i\over y_0}\dot z^2_0 = 0, \eqn\sanyon
$$
which is the geodesic equation on H with the Poincar\'e
metric\findref\hadamard.
The solutions to \sanyon~ are give by,
\foot{The second solution is in fact obtained by taking a
	proper limit of the first solution. The first solution is
	always transformed into the second one by an appropriate
	M\"obius transformation.}
$$
   z_0(t)=c_1{\sinh{X_0}+i \over{\cosh{X_0}}}+c_2,
  ~~~{\rm and}~~~~ ie^{X_0}+c_2, \eqn\sango
$$
where
$$
       X_0\equiv \omega(t+t_0),~~~~~~~~~~~
       c_1,\,c_2,\,\omega,\,t_0 \in {\rm I}\!{\rm R}. \eqn\sanroku
$$
	A classical motion is determined uniquely with the boundary
conditions which are the position and the velocity at the initial
point.  Thus the constants of the integration for the Euler-Lagrange
equation \sanni~or \sansan~  are  four real Grassmann even and
also four odd constants.    So expanding $z$ and $\theta$ in the
Grassmann odd constants, say, $\epsilon_1, \bar\epsilon_1,
\epsilon_2, \bar\epsilon_2$, we have a set of differential
equations for the coefficients of the Grassmann even functions.
However, it is actually not easy to solve those equations.
So instead of solving \sansan~ directly, we will take a roundabout.
	The Hamilton-Jacobi equation is given by
$$
{\partial S\over{\partial t}}
	+{1\over2m}\,g^{BA}\,{\partial S\over{\partial q^A}}
	\,{\partial S\over{\partial q^B}} =0 ~.\eqn\sannana
$$
Since the action which the classical solutions are plugged into
satisfies the above equation \sannana, we express $S$ as
$$
  S(q_1,q_2;t_2-t_1)={m\over2}\int^{t_2}_{t_1}dt~\dot q^A(t)\,
    g_{AB}(q(t))\,\dot q^B(t)~~,\eqn\sanhachi
$$
where $q^A(t)$ is a solution of the geodesic equation \sanni~
connecting the initial point $q_1=q(t_1)$ and the final point
$q_2=q(t_2)$.  It can be easily shown that the integrand is
time-independent and its body part is non-negative.  Taking them
into account, we set
$$
   \dot q^A(t)\,g_{AB}(t)\,\dot q^B(t)=({\rm const.})
    \equiv\omega^2~,\eqn\sankyu
$$
and define a superanalog of the hyperbolic distance,
$$
d(q_1,q_2)=\int^{t_2}_{t_1}dt~
	\sqrt{\dot q^A(t)\,g_{AB}(t)\,\dot q^B(t)}
    	=|\omega|(t_2-t_1)~.\eqn\yonjuu
$$
	From \sanhachi, \sankyu~and \yonjuu, we get
$$
      S(q_1,q_2;t)= {m\over2}{[d(q_1,q_2)]^2\over t}~~.
	\eqn\yonichi
$$
Note that the hyperbolic distance $d_0(q_1,q_2)$ between
(~{\tenrm (}$q_1${\tenrm )}$_0~) = (z_0,\bar z_0)$ and
(~{\tenrm (}$q_2${\tenrm )}$_0~) = (w_0,\bar w_0)$, which should be
the body part of $d(q_1,q_2)$, is given by
$$
   \cosh{d_0}=
   1+ {|z_0-w_0|^2\over{2\,{\rm Im}\,z_0\,{\rm Im}\,w_0}}
   \equiv 1+{1\over2}R_0~,  \eqn\yonni
$$
which is PSL(2,I$\!$R)-invariant.  And hence $d(q_1,q_2)$ should be
symmetric under the exchange of $q_1$ and $q_2$ and \SPL-invariant.
There exist two basic functions on SH$\times$SH with such
properties\findref\baranov,
$$
\eqalignno{
   R(q_1,q_2) &= {{|z_1-z_2-\theta_1\theta_2|}^2\over
   {Y_{(1)}Y_{(2)}}}~,&\eqnalign{\yonsan}\cr
   r(q_1,q_2)&= \left\{
  {2\theta_1\bar{\theta}_1+i(\theta_2-i\bar{\theta}_2)
  (\theta_1+i\bar{\theta}_1) \over {4Y_{(1)}} }+(1\leftrightarrow 2)
   \right\}& \cr
   {}& \qquad\qquad+{(\theta_2+i\bar{\theta}_2)
  (\theta_1+i\bar{\theta}_1)
  {\rm Re}(z_1-z_2-\theta_1\theta_2)\over{4Y_{(1)}Y_{(2)}}}~,
	&\eqnalign{\yonyon}\cr}
$$
where $Y_{(i)}={\rm Im}~z_i+{1\over2}\theta_i\bar{\theta_i}$ for
$i=1,2$. Here $R$ is the superanalog of $R_0$ in \yonni~ and $r$ is
nilpotent, and hence we can expect that $d(q_1,q_2)$ takes in general
the following form,
$$
	\cosh d =f(R)+ k(R)\,r~. \eqn\yongo
$$
The Hamilton-Jacobi equation leads to the differential equations for
the unknown functions $f$ and $k$,
and we find that the \lq\lq super''
hyperbolic distance $d(q_1,q_2)$ is given by\findref\ourfourth,
$$
   \cosh[d(q_1,q_2)] = 1+{1\over2}R(q_1,q_2) + k(R)\,r(q_1,q_2),
  \eqn\yonhachi
$$
where
$$
  \eqalign{k(R)&=\cosh l-1-\sinh l~\coth{l\over2a}~~,\cr
   l&=l(q_1,q_2)=\cosh^{-1}{\Big(}1+{1\over2}R{\Big)}~~~.\cr}
    \eqn\yonkyu
$$

The next step is to solve $q_1\equiv q$ in terms of $q_2$ and
its canonical conjugated quantity, say, $p^{(2)}$.
This can be done by solving the following algebraic equations
with respect to $q$,
$$
    {\partial S\over{\partial q_2^A}} =-p^{(2)}_A,  \eqn\gojuu
$$
where $q_2$'s and $p^{(2)}$'s actually correspond to the constants
of integration for the differential equations \sansan.
	The calculation is cumbersome but rather straightforward and
we obtain the solution of the Euler-Lagrange equations \sansan,
$(z^{\I}(t),\theta^{\I}(t))$,
$(z^{\Ini}(t),\theta^{\Ini}(t))$ or
$(z^{\Isan}(t),\theta^{\Isan}(t))$\findref\ourfourth\ourfifth\endrefs,
$$
\eqalignno{
z^{\I}(t) &=  \Big[ c_1-{2\over{\chx}}
     \{i\xi_1\xi_2e^{-{X\over a}}- i\xi_3\xi_4e^{X\over a}{}&{}\cr
     {}&{}\qquad\quad-\xi_1\xi_4e^{(1-{1\over a})X}
        +\xi_2\xi_3e^{({1\over a}-1)X}\}\Big]{\sinh X+i \over \chx}
        +c_2~,{}&{}\cr
   \theta^{\I}(t)&=\left({\sinh X+i \over\chx}+1\right)
      \{\xi_1e^{-{X\over a}}-i\xi_2e^{-X}+i\xi_3e^{({1\over a}-1)X}+
      \xi_4\},&\eqnalign{\goni}\cr
  z^{\Ini}(t) &= ie^X+c_2{}&{}\cr
     {}&{}\quad+i\xi_1\xi_2 e^{(1-{1\over a})X}
      -i\xi_3\xi_4 e^{(1+{1\over a})X}-\xi_1\xi_4e^{(2-{1\over a})X}+
       \xi_2\xi_3e^{{X\over a}},{}&{}\cr
  \theta^{\Ini}(t)&= i\xi_1e^{(1-{1\over a})X}+\xi_2
	   -\xi_3e^{{X\over a}} +i\xi_4e^X~,&\eqnalign{\gosan}\cr
   z^{\Isan}(t)&= ic_1+c_2-2ac_1\omega_st {}&{}\cr
       {}&{}\quad -\omega_s \{2ia^2c_1\omega_s -a\ebni \eichi
        +(1-a)\eni \ebichi \}t^2
        -{1\over3}\eichi \ebichi\omega_s t^3~,{}&{}\cr
     \theta^{\Isan}(t)&= \eni +\eichi t +
        \Big[ \{ia\eichi + (1-a)\ebichi \}\omega_s -
       {1-a\over2ac_1}\eichi\ebichi\eni \Big] t^2~,
	&\eqnalign{\goyon}\cr}
$$
where
$$
   \eqalign{
 {}&{}X\equiv \omega(t+t_0)~,\cr
 {}&{}\omega,t_0,c_1,c_2:~{\rm real~Grassmann~even~constants},~
        c_1>0, \cr
 {}&{}\xi_k~(k=1,2,3,4):~{\rm Grassmann~odd~constants~with~}
    \bar\xi_k=i\xi_k,\cr
 {}&{}\eichi, \eni :~{\rm complex~Grassmann~odd~constants},\cr
 {}&{}\omega_s:~{\rm Grassmann~even~constant~with~no~body~part}.
   \cr}\eqn\goroku
$$
The first ($z^{\I},\theta^{\I}$)  and the second
($z^{\Ini},\theta^{\Ini}$)  solutions correspond to the first and the
second solutions in \sango, respectively and the third one
($z^{\Isan},\theta^{\Isan}$) corresponds to the solution with
$\omega=0$  in \sango. And actually
($z^{\Ini},\theta^{\Ini}$)  is obtained by taking a proper limit of
($z^{\I},\theta^{\I}$).

	Now we examine the classical motion on the SRS.
Since we have obtained the classical paths \goni, \gosan\ and
\goyon\ on the covering space SH of the SRS, we can deduce the
classical motion on the SRS through projecting the paths on SH onto
the fundamental domain SH/\sgamma. We study closed orbits on the SRS
at first. A path $Z(t)=(z(t),\theta(t))$ on SH gives a closed loop on
the SRS if it satisfies the condition that there exist such an
element $k\ne\!1$ in \sgamma\  and a time interval $T$ that
$$
         Z(t+T)=k(Z(t))~.   \eqn\gohachi
$$
Since $k$ is characterized by the two fixed points, $(u,\mu$) and
$(v,\nu)$, the sign factor $\chi(k)$ and the norm function $N_k$
(see Sect.2),
the above condition gives a necessary condition,
$$
  \eqalign{
    {{z(t+T)-u-\theta(t+T)\mu}\over{z(t+T)-v-\theta(t+T)\nu}}&=
      N_k{{z(t)-u-\theta(t)\mu}\over{z(t)-v-\theta(t)\nu}}~, \cr
    {}&{}\cr
   {{\theta(t+T)+{\nu-\mu \over{u-v}}z(t+T)+{v\mu-u\nu \over{u-v}}}
      \over{z(t+T)-v-\theta(t+T)\nu}}&=
    \chi(k)\,N^{1/2}_k{{\theta(t)+{\nu-\mu \over{u-v}}z(t)+{v\mu-u\nu
    \over{u-v}}}\over{z(t)-v-\theta(t)\nu}}~.\cr}  \eqn\gokyu
$$
We find that the classical motions $Z^{\Ini}(t)$  \gosan\ and
$Z^{\Isan}(t)$ \goyon\ do not satisfy the above condition
\gokyu\ and only the motions $Z^{\I}(t)$ \goni\ with the
parameters having values,
\foot{
     We may interchange $(u,\mu)$ and $(v,\nu)$ in the
following equations, which corresponds to changing the sign of
$\omega$, or the direction of motion.} $\>\!\!\!\!\!$'
\foot{
      The last two conditions $\xi_1=\xi_3=0$ are not
necessary for $Z^{\I}(t)$ to satisfy the condition \gokyu~ if
$a=2$ and $n>0$, however, they are necessary to satisfy the
condition \gohachi.}
$$
   c_1={v-u\over2},~~c_2={u+v\over2},~~\xi_2={\nu\over2},
   ~~\xi_4={\mu\over2},~~\xi_1=\xi_3=0, \eqn\rokujuu
$$
satisfy the original condition \gohachi\ and the time interval $T$ is
$$
            T={\log N_k \over {\omega}}.  \eqn\rokuichi
$$
The path $Z^{\I}(t)$ with \rokujuu,  which we denote $Z_k(t)$
associated with the element $k$, is the geodesic
curve connecting the two fixed points of the element $k\ne\!1$ in
\sgamma;
$$
   \eqalign{
     Z_k(t \rightarrow +\infty)&{}\longrightarrow (v,\nu),\cr
     Z_k(t \rightarrow -\infty)&{}\longrightarrow (u,\mu),~~~
             \omega>0.\cr} \eqn\rokuni
$$
A segment $[Z_k(t),Z_k(t+T)]$ of the geodesic curve becomes a
closed loop on the SRS and the length of the loop $l(k)$ is given by
$$
  l(k) \equiv d(Z_k(t),Z_k(t+T))=d(Z_k(t),k(Z_k(t)))=\log N_k,
   \eqn\rokusan
$$
which in fact depends only on the element $k\in$\sgamma. Equation
\rokusan\ yields
$$
  l(k^{n})=\vert n\vert\, l(k). \eqn\rokugo
$$
The geodesic segment $[Z_k(t),Z_k(t+nT)]$ becomes a closed loop
lying $|n|$-fold exactly on the closed loop coming from the segment
$[Z_k(t),Z_k(t+T)]$. So $[Z_k(t),\break Z_k(t+nT)]$
and $[Z_k(t),Z_k(t+T)]$ determine the same primitive periodic orbit,
and we conclude that two elements
$k^m$ and $k^n \ne\!1 (m,n:{\rm integers})$ in \sgamma~ are
associated with the same primitive periodic orbit on the SRS.
Furthermore, due to \SPL-invariance of $d(q_1,q_2)$,
we get
$$
   l(k)=d(gZ_k(t),gZ_k(t+T))=d(gZ_k(t),g\,k\,g^{-1}(gZ_k(t))),
    ~~~~g\in{\rm S}\Gamma. \eqn\rokunana
$$
This implies that $gZ_k(t)$ is the geodesic curve connecting the two
fixed points of the element $g\,k\,g^{-1}\in$\sgamma.  Since
$gZ_k(t)$ and $Z_k(t)$ become the same trajectory on the SRS, we
conclude that every geodesic curve connecting the fixed points of
each element of Conj(\sgamma) becomes the
same orbit on the SRS.   Thus we find that each pair $(p,p^{-1})\in
{\rm Prim}({\rm S}\Gamma)$ is associated with a primitive periodic
orbit on the SRS and its length is given by $$
   l(p)=\log N_p=\log N_{p^{-1}_{}}~~,  \eqn\rokuhachi
$$
where $N_p$  is the norm function associated with $p$.
Conversely any periodic orbit can be lifted to a geodesic segment
$[Z(t),k(Z(t))]$ on SH with some element $k\ne\!1$ in \sgamma.
Since there exists a unique geodesic curve connecting the two points
$Z(t)$ and $k(Z(t))=Z(t+T)$, the geodesic curve is in fact a
solution $Z^{\I}(t)$ connecting the two fixed points of $k$.  Then
we conclude that there exists a one-to-one correspondence between
primitive periodic orbits on the SRS and pairs
of inconjugate primitive elements $(p,p^{-1})$.  Any geodesic curve
$Z^{\I}(t)$ not connecting two fixed points of any element in
\sgamma~ becomes a nonperiodic orbit on the SRS and such geodesic
curves are {\em dense} on SH.  Hence the classical motion on the
SRS is chaotic.

\noindent Signals of Chaos:

\item{1.}
the lagrangian \sanichi\ is \SPL-invariant, however, after
projecting out onto the SRS, we find that the symmetry generators on
SH become no longer those on the SRS and only two Grassmann even
quantities are conserved ones, which are the Hamiltonian $H$ and a
nilpotent quantities $H^{(2)}$ essentially corresponding to
$E^{\theta} E^{\bar{\theta}}$.
\foot{The explicit form of $H^{(2)}$ is $$H^{(2)}=
	Y(\theta\bar\theta p_zp_{\bar z}-\theta p_zp_{\bar\theta}-
	\bar\theta p_{\bar z}p_{\theta}-p_{\theta}p_{\bar\theta})~~,$$
	where $p_A(A=z,\bar{z},\theta,\bar\theta)$ denotes the
	canonical momentum.
	}
The fact that there are two kind of conserved
quantities has been already presented in constructing the
Lagrangian which consists of two \SPL-invariant pieces.
However, the dimension of the hyper surface determined by $H=E$ and
$H^{(2)}=E^{(2)}$ ($E,E^{(2)}$: constants) in the total super space
becomes less by one bosonic degree than that of total space
according to the (super) implicit function theorem\findref\berezin.
\item{2.}
	we will study the Anosov property\findref\anosov\ which
describes the behavior of the initially neighboring trajectories at
large times and is suitable to study the strongly chaotic
systems\findref\sinai.
Let us take two geodesic curves $Z^{\I}(t)$ $(\omega >0,t_0=0)$
with the conditions \rokujuu~ and another one with
$$
  c_1={v+\delta v-u\over2},~~c_2={u+v+\delta v\over2},~~
   \xi_2={\nu+\delta\nu \over2},
   ~~\xi_4={\mu\over2},~~\xi_1=\xi_3=0. \eqn\rokukyu
$$
These two trajectories start from the same point $(u,\mu)$ at
$t\rightarrow -\infty$, however, arrive at slightly different
points $(v,\nu)$ and $(v+\delta v,\nu+\delta\nu)$ when
$t\rightarrow \infty$. At $t=0$ the value of separation is
$$
    d_{t=0} \sim \delta v + {\mu+\nu \over2}\delta\mu,
     \eqn\nanajuu
$$
However as $t\rightarrow \infty$ the trajectories separate
exponentially,
$$
   d_{t\rightarrow\infty}\sim (\delta v+\nu\delta\nu)e^{\omega t}.
    \eqn\nanaichi
$$
The velocity $\omega$ is the Liapunov exponent.  This implies that
trajectories are unstable, which is characteristic of classical
chaos.
\item{3.}
	we comment on the Kolmogorov-Sinai entropy
$h$\findref\pesin\ which,  roughly speaking, measures unpredictability
of the motions.  This number comes out in the asymptotic formula for
the counting function of primitive orbits of period $T(p)\leq T$,
$$
   \#\{p,\ T(p)\leq T\} \sim {e^{hT}\over{hT}},~~~~~~
   T\rightarrow +\infty, \eqn\nanani
$$
which indicates the exponential proliferation of the periodic orbits.
	From \rokuichi\ and \rokuhachi, this formula yields
$$
\#\{p,l(p)\leq x\}\sim {e^{\alpha x}\over{\alpha x}},~~x\rightarrow
\infty~~{\rm with}~~\alpha={h\over\omega}. \eqn\nanasan
$$

\noindent The asymptotic formula \nanasan\ will be discussed in the
quantum  mechanical framework in Sect. 6.

	\chapter{Quantum Mechanics}

	In this section we develop the quantum theory for a particle
moving on the SRS.  As  was the conventional Hadamard model, we
cannot {\it quantize} the classical motion on the SRS because of its
ergodicity.  However there exists a well-defined quantum mechanics
on the SRS where the Laplace-Beltrami operator acts as the
quantum Hamiltonian. To construct the quantum mechanics on the SRS,
we firstly develop the quantum mechanics on SH where the quantum
Hamiltonian is also the Laplace-Beltrami operator.  The quantum
motion on the SRS is obtained by imposing the {\em periodic}
boundary conditions upon the motion on SH just as the case of the
classical motion.  The quantum motion on SH is also expected to be
integrable.  In fact, as we will see, we can solve the
Schr\"odinger equation explicitly and obtain the exact energy
spectrum for the quantum motion on SH.  A wave function on the
SRS will be obtained by {\it folding} that on SH, however the function
is quite complicated.

	First we give the quantum Hamiltonian.
Our Lagrangian \sanichi\ is nonlinear in a sense that
$g_{AB}$ are functions of supercoordinates.  Omote and
Sato\findref\omote\ developed a procedure to construct Hamiltonian for
a system with a (purely bosonic) nonlinear  Lagrangian of a form
$L_B={1\over2} g_{ij}(q)\,\dot q^i \,\dot q^j$  with having considered
a symmetry as a guiding principle (see also\findref\kawai).  We can
follow their arguments paying attention to sign factors.  We find the
quantum Hamiltonian on SH,
$$
 H_Q={(-)^A\over 2m}g^{-1/4}\,p_A\,g^{1/2}g^{AB}\,p_B\,g^{-1/4}~,
    \eqn\nanayon
$$
where
$$
     g \equiv | {\rm sdet}\,g_{AB} | = (4a^2Y^2)^{-1}~,\eqn\nanago
$$
with the canonical commutation relations,
$$
    [\,p_A\,,\,q^B\,]_{\pm}= -i\hbar\,\delta_A^{\ B}~.\eqn\nanaroku
$$
The Hamiltonian $H_Q$ is also \SPL-invariant and hence is the
Hamiltonian on SRS.

In the $q$-representation the coordinates $q^A$ and the momenta
$p_A$ are given by
$$
  \eqalign{ q^A&=q^A~, \cr
   p_A&=-i\hbar g^{-1/4}{\partial\over{\partial q^A}}\,g^{1/4}
   \equiv -i\hbar g^{-1/4}\,\dela\,g^{1/4}~, \cr}\eqn\nananana
$$
so that $H_Q$ is a super Laplace-Beltrami operator,
$$
  \eqalign{H_Q&=-{\hbar^2\over2m}(-)^A g^{-1/2}
      \dela (g^{1/2}g^{AB}\delb)\cr
   {}&=-{\hbar^2\over2m}\left[(2YD{\overline D})^2+{1-a\over a}
    (2YD{\overline D})\right] \equiv
     {\hbar^2\over2m}\lap~,\cr} \eqn\nanahachi
$$
where $D$ is given in \go~ and $\overline D$ is its complex
conjugate. The Hilbert space {\cal H} of our model is the space of
superfunctions on SH with the inner product
$$
  \VEV{\Psi_1{\big|}\,\Psi_2} \equiv \int d^4q\,g^{1/2}(q)
   \VEV{\Psi_1{\big|}q}
   \VEV{q{\big|}\Psi_2}
   , \eqn\nanakyu
$$
and $H_Q$ is hermitian with respect to this product.

	We will study the spectral properties of $H_Q$ on SH.
In order to do that we examine the eigenvalue problem of the operator
$\hako$,
$$
     \hako \equiv 2YD\overline D~. \eqn\hachijuu
$$
	The Grassmann even (odd) eigenfunction $e_{\Lambda}
(\psi_\Lambda)$, with eigenvalue $\lambda^B (\lambda^F)$,
$$
\eqalign{
     \hako \,e_{\Lambda} &= \lambda^B\, e_{\Lambda}~,\cr
     \hako \,\psi_{\Lambda} &= \lambda^F\, \psi_{\Lambda}~,\cr}
	\eqn\hachiichi
$$
may be expanded as
$$
\eqalign{
e_{\Lambda}& = A_{\Lambda}+\theta\bar{\theta}B_{\Lambda}~,\cr
\psi_{\Lambda}&={1\over{\sqrt y}}(\theta\rho_{\Lambda}+\bar\theta
	 \varphi_{\Lambda})~,\cr} \eqn\hachini
$$
where $\{A_{\Lambda},B_{\Lambda}\}$ and
$\{\rho_\Lambda,\varphi_\Lambda\}$ are functions of Grassmann
even coordinates $z=x+iy$ and $\bar z$.  Eqs. \hachiichi~and
\hachini~yield
$$
\eqalign{{}&B_{\Lambda}={\lambda^B\over{2y}}A_{\Lambda}~, \cr
	{}&{\Bigl\{}y^2({\delx}^2+{\dely}^2)
	-\lambda^B(\lambda^B-1){\Bigr\}} A_{\Lambda}=0~.\cr
    {}& \varphi_{\Lambda}=-{2y\sqrt y \over{\lambda^F}} \delzb
   \left(\rho_{\Lambda}\over{\sqrt y}\right)~, \cr
   {}& \Bigl\{y^2(\delx^2+\dely^2)-iy\delx
   -((\lambda^F)^2-{1\over4})\Bigr\} \rho_{\Lambda}=0~,\cr}\eqn\hachisan
$$
and the above differential equations can be solved as
\foot{When $k=0$, $A_\Lambda\sim y^\lambda$ or $y^{1-\lambda}$ and
$\rho_\Lambda\sim y^{1/2\pm\lambda}$. Then they are unnormalizable on
SH.},
$$
\eqalign{
A_{\Lambda}&=C_{\lambda^B,k}e^{ikx}{\sqrt y}
     \, K_{\lambda^B-{1\over2}}(|k|y)~,\cr
\rho_{\Lambda}&=C_{\lambda^F,k}\,e^{ikx}\,
      W_{{\signk \over2},\lambda^F}(2|k|y)~,\cr}
	\eqn\hachiyon
$$
where $\signk \equiv {\rm sign}(k),\ k\ne 0$
and $K_\mu$ and $W_{\kappa,\mu}$ are a modified Bessel function and  a
Whittaker function, respectively.
	The normalization constants $C_{\lambda^B,k}$ and
$C_{\lambda^F,k}$ are determined as follows.
	Since SH is noncompact and the spectrum is continuous, the
normalization condition should be
$$
\eqalign{
\VEV{e_{\Lambda}{\Big|}e_{\Lambda'}} &\propto
  \delta(\lambda^B-(\lambda^B)')~,\cr
\VEV{\psi_{\Lambda}{\Big|}\psi_{\Lambda'}} &\propto
  \delta(\lambda^F-(\lambda^F)')~,\cr}\eqn\hachigo
$$
The above condition determines the regions of the
eigenvalues\findref\ourfifth,
$$
\eqalign{
\lambda^B &={1\over2}+ip~,\cr
\lambda^F &=c+ip~,\cr}\eqn\hachiroku
$$
where $p\in (-\infty,+\infty)$
and $\{c$:real constant,\ $|c|\leq{1\over2}\}$.
And the normalized eigenfunctions are given by,
$$
\eqalign{
e_{p,k}(Z)&=\left(2ia\sinh{\pi p}\over{\pi^3}\right)^{1/2}
    \left(1+{1+2ip\over{4y}}\theta\bar\theta\right)e^{ikx}{\sqrt y}
    \,K_{ip}(|k|y)~,\cr
\psi^c_{p,k}(Z)&= \left(
  {a\cos{[\pi(c+ip)]}\over{2\pi^2k(c+ip)^{\signk -1}}}\right)^{1/2}
    {1\over{\sqrt y}} e^{ikx}\cr
    {}&{}\quad\times\left\{\theta\,W_{{\signk\over2},c+ip}(2|k|y)
    +i(c+ip)^{\signk}
     \bar\theta\,W_{-{\signk \over2},c+ip}(2|k|y)\right\}~,\cr}
	\eqn\hachinana
$$
with
$$
\eqalign{
 \VEV{e_{q,l}{\Big|}e_{p,k}}&=\delta(p+q)\delta(k-l)~,\cr
 \VEV{\psi^c_{q,l}{\big|}\psi^c_{p,k}}&=\delta(k-l)\delta(p+q)~.\cr}
	\eqn\hachikyu
$$
The eigenvalues of $H_Q$ are
$$
\eqalign{
 E^B_{p,k}&={\hbar^2\over2m}\left\{\left({1-a\over2a}\right)^2
   +\left(p-{i\over2a}\right)^2 \right\}
    \equiv {\hbar^2\over2m}\gamma^B(p)~,\cr
 E_{p,k,c}^F&={\hbar^2\over{2m}}\left\{\left({1-a\over{2a}}\right)^2
  +\left(p-ic-i{1-a\over{2a}}\right)^2\right\} \equiv
    {\hbar^2\over{2m}}\gamma^F_c(p)~.\cr}\eqn\kyujuu
$$
Although $H_Q$ is a hermite operator, the eigenvalue is complex.
This is because the space of eigenstates contains isovectors
\findref\aoki\ as is seen in \hachikyu.
Notice that except when $c={1\over2}$ or ${a-2\over{2a}}$ with
$a\geq1$, the energy spectra of the Grassmann even states and the
odd ones do not coincide with each other,
$$
   \{E^B_{p,k}\}\ne\{E^F_{q,l,c}\}~,~~~~~~~~
   c\ne{1\over2}~{\rm and}~{a-2\over{2a}}~~(a\geq1)~. \eqn\hyaku
$$

	A set of eigenfunctions for each $c$,
$\{e_{p,k},\psi^c_{p,k}\}$, do satisfy the completeness relation,
$$
  \eqalign{\int^{\infty}_{-\infty}dpdk\,&
   [e_{p,k}(q_2)\,\bar e_{-p,k}(q_1)+\psi^c_{p,k}(q_2)\,
   \bar\psi^c_{-p,k}(q_1)\,] \cr
  {}&=[g(q_1)g(q_2)]^{-1/4}\ \delta(q_1-q_2)~.\cr }\eqn\hni
$$
For each $c$, we have a Hilbert space for the Grassmann odd states
${\cal H}^F_c$, and hence the total Hilbert space is
$$
   {\cal H}_c={\cal H}^B\oplus{\cal H}^F_c~~,\eqn\hsan
$$
where ${\cal H}^B$ is the Hilbert space for the Grassmann even
states.

	We now turn our attention to the eigenfunctions on the SRS or
F, a fundamental domain of \sgamma.  The \sgamma-invariance of the
eigenfunction $\Psi$ imposes the {\em periodic} boundary
condition.  The action of $g\in{\rm S}\Gamma$ on $\Psi$ reads
$$
\Psi'(q)=\left[g\Psi\right](q)=\Psi(g^{-1}q),~~q\in{\rm SH}.
\eqn\hyon $$
Hence, the periodicity is expressed as
$$
\Psi(g^{-1}q)=\Psi(q)~~~~{\rm for~~ all}~~g\in S\Gamma. \eqn\hgo
$$
The spectrum of the operator $\Delta_{SLB}$ on SH,
$\{\gamma^B(p)\}$ and $\{\gamma^F(p)\}$ in \kyujuu,
will become discrete on the SRS.  We then write the discrete
spectrum of $\Delta_{SLB}$ as
$$
 \eqalign{
&\{\gamma^B_n\}~~~ (n=0,1,2,\cdots)~~~{\rm for~ Grassmann~ even~
state},\cr &\{\gamma^F_n\}~~~ (n=0,1,2,\cdots)~~~{\rm for~
Grassmann~ odd~ state},\cr }\eqn\hroku
$$
and that of the operator $\hako$ in \hachijuu~ as
$$
 \eqalign{
&\{\lambda^B_n\}~~~ (n=0,1,2,\cdots)~~~{\rm for~ Grassmann~ even~
state},\cr &\{\lambda^F_n\}~~~ (n=0,1,2,\cdots)~~~{\rm for~
Grassmann~ odd~ state},\cr }\eqn\hnana
$$
where
$$
\gamma^{B(F)}_n=-(\lambda^{B(F)}_n)^2
	-({1-a\over a})\lambda^{B(F)}_n.  \eqn\hhachi
$$
However, because the periodic condition is complicated it is very
difficult to see the spectrum on the SRS explicitly.  We comment on
the ground state. The Grassmann even ground state is given by a
constant function.
It is a trivial periodic function with the normalization
$$
	 \int_F dV {\rm (const.)}<\infty,  \eqn\hkyu
$$
and has the energy $\lambda^B_0=0\ (\gamma^B_0=0)$.

	We now construct the kernel function on SH. The kernel
function is given by,
$$
 \eqalign{K(q_1,q_2;t)& \equiv \bra{q_2}e^{-{it\over\hbar}H_Q}
    \ket{q_1}~, \cr
     {}&=\int^{\infty}_{-\infty} dpdk
    \,{\Bigl\{}e^{-{it\over\hbar}E^B_{p,k}}\VEV{q_2{\big|}e_{p,k}}
    \VEV{e_{-p,k}{\big|}q_1}\cr
   {}&{}\qquad\qquad\qquad +
    e^{-{it\over\hbar}E^F_{p,k,c}}\VEV{q_2{\big|}\psi^c_{p,k}}
     \VEV{\psi^c_{-p,k}{\big|}q_1}{\Bigr\}}\cr
	{}& \equiv K(q_1,q_2|\tau)~,\cr}\eqn\hjuu
$$
where
$$
       \tau \equiv {i\hbar\over{2m}}t~. \eqn\hichiichi
$$
Plugging \hachinana\ into \hjuu, we
get\findrefs\ourfourth\ourfifth\endrefs,
$$
   K(q_1,q_2|\tau)=K^{(0)}(l;\tau)+r(q_1,q_2)\,K^{(1)}(l;\tau)~,
    \eqn\hichini
$$
where $l=l(q_1,q_2)$ is given in \yonkyu\ and,
$$
\eqalignno{
K^{(0)}(l;\tau)&={-2a\over{\pi\sqrt{2\pi \tau}}}\edelv
        \int^{\infty}_l db\,
   {{e^{-{b^2\over4\tau}}\sinh{b\over2a}}
     \over{(\cosh b-\cosh l)^{1/2}}}~,&\eqnalign{\hichisan}\cr
K^{(1)}(l;\tau)&={-2a\over{\pi\sqrt{2\pi \tau}}}\edelv
   \int^{\infty}_l db\,{1\over{(\cosh b-\cosh l)^{1/2}}}&\cr
   {}&{}\qquad\qquad \times{\Bigg[}(\cosh l-1){d\over db}
        {\bigg(}e^{-{b^2\over{4\tau}}}
    {\sinh{b\over2a}\over{\sinh b}}{\bigg)}&\cr
    {}&{}\qquad\qquad\qquad +e^{-{b^2\over{4\tau}}}
   \left({b\over2\tau}\cosh{b\over2a}+{a-1\over2a}\sinh{b\over2a}
   \right){\Bigg]}~.&\eqnalign{\hichiyon}\cr}
$$
So the time development for a wave function on SH is given by
$$
  \Psi(q,t)=\int\nolimits_{\rm SH}dV(q')\,
   K(q,q';t-t') \Psi(q',t')~. \eqn\hichigo
$$
As for the wave function on the SRS, we should have
$$
	\Psi_{SRS}(q,t)=\int_{SRS}dV(q')K_{SRS}(q,q';t-t')
	\Psi_{SRS}(q',t') \eqn\hichiroku
$$
The periodicity of $\Psi_{SRS}$ implies that a kernel $K_{SRS}$ on
the SRS is written as
$$
	K_{SRS}(q_1,q_2|\tau)=\sum_{g\in{\rm S}\Gamma}K(q_1,g(q_2)|\tau).
	\eqn\hichinana
$$

	\chapter{Trace Formula and Zeta Function}

	In the preceding sections, we have eventually solved the
quantum mechanics on SH and obtained the heat kernel on SH, which
yields that on the SRS.  Here in this section, we will concentrate on
the quantum energy spectrum on the SRS.  As we have seen that the
spectrum on the SRS was too complicated, it is quite difficult to
estimate it explicitly.  However, in the conventional case, that is,
in the case of a particle on the Riemann surface of genus $g\geq2$,
the energy spectrum is related to the length spectrum through the
Selberg trace formula\findref\balazs.  We may expect that a similar
relation will exist in our model.

	First we present a formula of supertrace of a function
$G_{\rm SRS}(q_1,q_2)$ on SH/\sgamma $\times$ SH/\sgamma\ which is
made out of a \SPL-invariant function $G(q_1,q_2)$ on SH$\times$SH;
$$
    G_{\rm SRS}(q_1,q_2)=\sum_{g\in{\rm S}\Gamma}\ G(q_1,g(q_2))~,
    \eqn\hichihachi
$$
with
$$
   G(q_1,q_2)=\Phi(l(q_1,q_2))+r(q_1,q_2)\,\Psi(l(q_1,q_2))~,
      \eqn\hichikyu
$$
where $\Phi$ and $\Psi$ are some functions and $l$ and $r$ are
given in \yonkyu\ and \yonyon, respectively.  We find that
the supertrace of $G_{\rm SRS}$ defined by,
$$
 {\rm str}\,G_{\rm SRS}=\int\nolimits_{\rm F}dV(q)\,
   G_{\rm SRS}(q,q)=\sgsum\,\int\nolimits_{\rm F}dV(q)\,G(q,g(q))~,
  \eqn\hnijuu
$$
where F is a fundamental domain of \sgamma, is calculated as
$$
    {\rm str}\,G_{\rm SRS}= Area({\rm SRS})~\Phi(0)+\sumprim
    \sum^{\infty}_{n=1}\int\nolimits_{{\rm F}_p}dV\ G(q,p^n(q)),
    \eqn\hniichi
$$
where use has been made of a formula,
$$
	\sum_{g\in{\rm S}\Gamma} f(g) = f(1) +
	\sumprim\sum^{\infty}_{n=1}\sum_{g\in{\rm S}\Gamma/Z(p)}
	f(gp^ng^{-1})~.\eqn\hniichia
$$
and F$_p$ is a fundamental domain for the centralizer of $p$, $Z(p)$,
$$
      {\rm F}_p=\bigcup\limits_{g\in{\rm S}\Gamma/Z(p)}~
       g^{-1}\,{\rm F}~~.\eqn\hnini
$$
We can assume that $p$ is a magnification with a matrix
$A_p={\rm diag}(\chi(p)N_p^{1/2},\break \chi(p)N_p^{-1/2},1)$
(see \ichiroku) and we can choose a convenient domain for F$_p$
\findref\baranov,
$$
   \int\nolimits_{{\rm F}_p}dV\Longrightarrow\int^{N_p}_1dy
      \int^{\infty}_{-\infty}dx\int d\theta d\bar\theta~
    {1\over{2ay+a\theta\bar\theta}}~~.\eqn\hnisan
$$
Then we finally get
$$
  \eqalign{ {\rm str}\,G_{\rm SRS}&={\pi(g-1)\over a}~\Phi(0)\cr
    {}&-\sumprim\sum^{\infty}_{n=1}\,
    {l(p)\over{2a\sqrt{\cosh{l(p^n)}-1}}}\cr
 {}&{}\quad\times{\bigg\{}\left(1-\chi(p^n)\cosh{l(p^n)\over2}\right)
   \int_{l(p^n)}^{\infty}ds\,
   {\sinh s\over{(\cosh b-\cosh{l(p^n)})^{1/2}}}~\Psi(s) \cr
   {}&{}\qquad\qquad+(1-\cosh{l(p^n)})\int_{l(p^n)}^{\infty}ds\,
    {1\over{(\cosh b-\cosh{l(p^n)})^{1/2}}}~
    {d\Phi(s)\over ds}{\bigg\}}~,
    \cr}   \eqn\hniyon
$$
where $g$ is the genus of the SRS and
$$
      l(p^n)=|n|l(p)=|n|\log N_p~~.\eqn\hnigo
$$

     Now we apply the above formula to the heat kernel on the SRS
\hichinana~ which can be written as
$$
 K_{\rm SRS}(q_1,q_2|\tau)=\sgsum\,\bra{q_1}e^{-\tau\lap}\ket{g(q_2)}~.
               \eqn\hniroku
$$
Then we get
$$
    {\rm str}\,K_{\rm SRS}={\rm str}\left(e^{-\tau\lap}\right)=
     \sum^{\infty}_{n=0}\,\left(e^{-\tau\gamma^B_n}-
       e^{-\tau\gamma^F_n}\right) ~.\eqn\hninana
$$
Plugging $K^{(0)}$ \hichisan\ and $K^{(1)}$ \hichiyon\ respectively
into $\Phi$ and $\Psi$ in \hniyon~ and integrating with respect to
$s$, we finally obtain a superanalog of Selberg trace
formula\findref\ourfifth,
$$
   \sum^{\infty}_{n=0}\,\left(e^{-\tau\gamma^B_n}-
       e^{-\tau\gamma^F_n}\right)=A(\tau)+\Theta(\tau)~,\eqn\hnihachi
$$
where
$$
\eqalignno{
  A(\tau)&={1\over\sqrt{\pi \tau}}(1-g)\edelv
	\int^{\infty}_0 db\ e^{-{b^2\over4\tau}}\,
	{\sinh{b\over2a}\over\sinh{b\over2}}~,&\eqnalign{\hnikyu}\cr
\Theta(\tau)&={1\over{4\sqrt{\pi\tau}}}\sumprim
	\sum^{\infty}_{n=1}{\rm str}{\Big(}K(p^n|a){\Big)}
	{l(p)\over\sinh{l(p^n)\over2}}e^{-{l^2(p^n)\over4\tau}-
	\left({1-a\over2a}\right)^2\tau},&\eqnalign{\hsanjuu}\cr
K(p|a)&={\rm diag}\left(e^{l(p)\over2a},e^{-{l(p)\over2a}},
	\chi(p)e^{{1-a\over2a}l(p)},\chi(p)e^{-{1-a\over2a}l(p)}\right)
   	~.&\eqnalign{\hsanichi}\cr}
$$
$A(\tau)$ is the contribution of the \lq\lq trivial motion'' on the
SRS (zero length term) and for $\tau\rightarrow 0$ it can be expanded
into a positive power series,
$$
  A(\tau)\sim -{Area({\rm SRS})\over{\pi}}
  (b_0+b_1\tau+b_2\tau^2+\cdots)~,
   \eqn\hsanni
$$
with
$$
  \eqalign{b_0&=1~,\cr b_1&={(a-1)(1-2a)\over6a^2}~,\cr
  b_2&={(a-1)(2a-1)(2a^2-2a+1)\over60a^4}~,\cr
  {}&{}\vdots\cr{}&{} \qquad .\cr} \eqn\hsansan
$$
This series approximates \lq\lq${\rm str}\left(e^{-\tau\lap}\right)$''
up to an exponentially small error.
On the other hand, $\Theta(\tau)$ is the contribution from the
periodic motions on the SRS and consistent with the semiclassical
approximation\findref\ourthird.

	Let us introduce a zeta function $Z(s|a)$
\findref\ourfourth\ with  one parameter $a$ associated with our model.
The function is defined by,
$$
  Z(s|a)\equiv\proprim
   \prod_{n=0}^{\infty}\,{\rm sdet}\,\left(1-K(p|a)
    e^{-(s+n)l(p)}\right)~,\eqn\hsanyon
$$
and we see that the zeta function is related to the trace
formula\findref\ourfifth,
$$
{d\over ds}\log Z(s|a)=(2s-1)\int^{\infty}_0dt
  \,e^{-\left(s-{1\over2}\right)^2t+\left({1-a\over2a}\right)^2t}
\,\Theta(t)~.\eqn\hsango
$$

	We point out that the zero-points and the poles of
$Z(s|a)$ give directly the eigenvalues of $\lap$ (energy spectrum)
on the SRS. More precisely, the zero-points give the eigenvalues of
the Grassmann even functions and the poles give those of the odd
functions. Using the trace formula, we find
$$
  \eqalign{{}&{}{d\over ds}\log{Z(s|a)}\cr
   {}&{}\quad=(2s-1)\int^{\infty}_0dt\,
  e^{-\left(s-{1\over2}\right)^2t+\left({1-a\over2a}\right)^2t}
  \left\{\sum^{\infty}_{n=0}
   (e^{-t\gamma^B_n}-e^{-t\gamma^F_n})-A(t)\right\}\cr
  {}&{}\quad=(2s-1)\sum^{\infty}_{n=0}\left[
  {1\over{\left(s-{1\over2}\right)^2-\left({1-a\over2a}\right)^2+
   \gamma^B_n}}-
    {1\over{\left(s-{1\over2}\right)^2-\left({1-a\over2a}\right)^2+
   \gamma^F_n}}\right]\cr
   {}&{}\qquad\qquad\qquad+2(g-1)\sum^{\infty}_{n=0}\,
   \left({1\over s+n-{1\over2a}}-{1\over s+n+{1\over2a}}\right)~.
   \cr}\eqn\hyonjuu
$$
Note that the last term in \hyonjuu~ becomes a finite sum when
$|a|^{-1}=m$ ($m$: positive integers),
$$
    2(g-1)~{\rm sign}(a)~\sum^{m-1}_{n=0}{1\over s+n-{m\over2}}~~.
   \eqn\hyonichi
$$
This formula implies that $Z(s|a)$ has a meromorphic continuation
onto the whole complex plane C$\!\!\!${\elevenrm l}~.
The zero-points (poles) of order $1$ exist at
$$
    s={1\over2}\pm\sqrt{\left({1-a\over2a}\right)^2-
   \gamma_n^{B\,(F)}}~~~.\eqn\hyonni
$$
Other trivial zero-points (ZP) and poles (P) exist respectively at,

\item{I.}$a^{-1}\ne m$ ($m$:positive integers);
$$
   \eqalign{{\rm ZP}{}&{}:\quad s=-n+{1\over2a}~~,\cr
    {\rm P}{}&{}:\quad s=-n-{1\over2a}~~,
     ~~ (n=0,1,2,\cdots)~,\qquad\cr}\eqn\hyonsan
$$
\item{I$\!$I.} $a^{-1}=m$;
$$
   \eqalign{
    {\rm ZP}{}&{}:\quad s=-n+{m\over2}~,\cr
     {\rm P}{}&{}:~\quad ~~{\rm none}~~,
     \qquad\qquad(n=0,1,\cdots,m-1)~,\cr}\eqn\hyonyon
$$
\item{I$\!$I$\!$I.} $a^{-1}=-m$;
$$
   \eqalign{
    {\rm ZP}{}&{}:\quad~~~{\rm none}~~,\cr
           {\rm P}{}&{}:~\quad s=-n-{m\over2}~,
    \quad (n=0,1,\cdots,m-1)~,\cr}\eqn\hyongo
$$

\noindent where the order of each ZP and P is $2(g-1)$ .

	\chapter{Classical Chaos from the Trace Formula}

	Our aim in this section is to discuss the exponential growth
of the counting function $\Pi(x)$ for the lengths of primitive
periodic orbits in \nanasan,
$$
\eqalign{
	\Pi(x)&=\#\{p,l(p)\leq x\}\cr
      &\sim {e^{\alpha x}\over{\alpha x}}~,\qquad\qquad\qquad
	{\rm for}~x\rightarrow\infty.\cr }\eqn\hgojuu
$$
 For this purpose, we prefer the general Selberg supertrace formula,
established in Refs.\findrefs\baranov\grosche\endrefs.
Specifically, if $h$ is a test function with the properties

\item{({\rm i})} $h({1\over2}+ip)
	\in \cplain\,^\infty({\rm I\!R})$
\item{({\rm ii})} $h({1\over2}+ip)\sim O({1\over p^2})
	~~{\rm for}~p\rightarrow\pm\infty$
\item{({\rm iii})} $h({1\over2}+ip)$ is holomorphic in the strip
	$|{\rm Im}\,p|\leq1+\epsilon, \epsilon>0,$

\noindent then the supertrace formula on the SRS with genus $g\geq2$
is given by
\foot{The trace formula reduces to \hnihachi\ if one chooses the test
function $h$ as $h(\lambda^{B(F)}_n) = e^{-\gamma^{B(F)}_n\tau}$, where
	$ {\gamma^{B(F)}_n = -(\lambda^{B(F)}_n)^2
	- {1-a\over a}\lambda^{B(F)}_n}$.}
\findref\baranov,
$$
\eqalign{
\sum^{\infty}_{n=0} & \left[h(\lambda^B_n)-h(\lambda^F_n)\right]\cr
	{}&=(1-g)\int^{\infty}_{0}{g(u)-g(-u)\over{\sinh{u\over2}}}du\cr
	{}&{}\qquad+\sumprim\sum^{\infty}_{n=1}{l(p)\over{2\sinh{nl(p)\over2}}}
	\Bigg[ g(nl(p))+g(-nl(p))\cr
	{}&{}\qquad\qquad-\chi^n(p) \left( g(nl(p))\,
	e^{-nl(p)\over2} + g(-nl(p))\,e^{nl(p)\over2}
	\right)\Bigg]~,\cr }\eqn\hgoichi
$$
where $\{\lambda^{B(F)}_n\}$ is the spectrum of
the operator $\hako$ (see \hnana) and
$$
	g(u)={1\over 2\pi}\int^{\infty}_{-\infty}dp~
	e^{-iup}h({1\over 2}+ip). \eqn\hgoni
$$
To get the information for $\Pi(x)$, it is convenient to choose a test
function $h$ so that the term proportional to $\chi^n(p)$ in
\hgoichi\ cancel, i.e.,
$$
	g(nl(p))e^{-nl(p)\over 2}+g(-nl(p))e^{nl(p)\over 2}=0~.
	\eqn\hgosan
$$
We take a function $({\rm Re}~s>1,{\rm Re}~\sigma>1)$
$$
	h(\lambda)=2\lambda({1\over s^2-\lambda^2}
	-{1\over{\sigma^2-\lambda^2}})~, \eqn\hgoyon
$$
with the Fourier transform $g(u)$ given by\findref\grosche,
$$
	g(u)={\rm sign}(u)e^{u\over 2}(e^{-s|u|}-e^{-\sigma|u|})~.
	\eqn\hgogo
$$
Thus only $\chi(p)$-independent term remains in the supertrace formula.
Plugging \hgoyon\ and \hgogo\ into \hgoichi, we have
$$
\eqalign{
	2\sum^{\infty}_{n=0} & \left\{{\lBn\over s^2-(\lBn)^2}-
	{\lFn\over{s^2-(\lFn)^2}}-{\lBn\over{\sigma^2-(\lBn)^2}}
	+{\lFn\over{\sigma^2-(\lFn)^2}}\right\}\cr
	&=4(g-1)\{\Psi(1+s)+\Psi(s)-\Psi(1+\sigma)-\Psi(\sigma)\}\cr
	{}&{}\qquad+\sumprim l(p)\left({e^{-sl(p)}\over1-e^{-sl(p)}}
	-{e^{-\sigma l(p)}\over1-e^{-\sigma l(p)}}\right)\cr}
\eqn\hgoroku
$$
where $\Psi(z)=\Gamma'(z)/\Gamma(z)$.  Defining
$$
	F(s)=\sumprim l(p){e^{-sl(p)}\over 1-e^{-sl(p)}}~,\eqn\hgonana
$$
and using a formula,
$$
	\Psi(z)=-\gamma-\sum^{\infty}_{n=0}\left({1\over z+n}
	-{1\over{n+1}}\right)~, \eqn\hgohachi
$$
we obtain
$$
\eqalign{
F(s)&-F(\sigma)\cr
 = & 4(g-1)\{\sum^{\infty}_{n=0}{1\over 1+s+n}
	+\sum^{\infty}_{n=0}{1\over s+n}
	-\sum^{\infty}_{n=0}{1\over 1+\sigma+n}
	-\sum^{\infty}_{n=0}{1\over\sigma+n}\}\cr
 {}&{}+\sum^{\infty}_{n=0}\{{1\over s-\lBn}-{1\over s+\lBn}
	-{1\over s-\lFn}+{1\over s+\lFn}\}\cr
 {}&{} -\sum^{\infty}_{n=0}\{{1\over \sigma-\lBn}-{1\over\sigma+\lBn}
	-{1\over \sigma-\lFn}+{1\over \sigma+\lFn}\}.\cr
	}\eqn\hgokyu
$$
Now we can read the analytic properties of $F(s)$ from the above trace
formula
\foot{We have assumed the convergence of the series\hgokyu.}:

\item{({\rm i})} $F(s)$ has a meromorphic continuation onto the
whole plane $\cplain$ ,
\item{({\rm ii})} $F(s)$ has poles in the following points
$$
\eqalign{
s&=0,\qquad\qquad\qquad\qquad~~~~  {\rm residue}~~~~ 4(g-1), \cr
s&=-n~~(n=1,2,\cdots),~~~~~~   {\rm residue}~~~~ 8(g-1), \cr
s&=\pm\lBn~~{\rm and}~~ \lBn\neq\lFn,~~~{\rm residue}~~~\pm1, \cr
s&=\pm\lFn~~{\rm and}~~ \lBn\neq\lFn,~~~{\rm residue}~~~\mp1. \cr
}\eqn\hrokujuu
$$

     Now let us return to the asymptotic formula of $\Pi(x)$ in
\hgojuu. We will see that the exponential coefficient $\alpha$ in the
equation is determined by the pole of $F(s)$ corresponding to the real
eigenvalue of $\hako$. First we consider the series,
$$
	F(k)=\sumprim l(p){e^{-kl(p)}\over 1-e^{kl(p)}}
	~~~~{\rm for~real~positive}~k~. \eqn\hrokuichi
$$
The convergence of the above series is under control of the balance
between the proliferation of the number of the paths and the
exponential damping factor $e^{-kl(p)}$.  When the variable
$k$ moves to the small direction, the singularity of $F(k)$, which is
connected with the largest real eigenvalue through the trace formula
\hgokyu, appears on the real positive axis.  On the other hand, we expect
that the eigenvalues of $\hako$ locate on (see \hachiroku),
$$
\eqalign{
	{\rm Re}~(\lBn)&={1\over2},\cr
	{\rm Re}~(\lFn)&= c~~~(|c|\le{1\over2}).\cr
	}\eqn\hrokuni
$$
Thus the largest positive eigenvalue $\lambda_{max}$ should exsist and it
takes the value ${1\over2}$ or $c (>0)$ if we assume its existance and ignore
so called {\it small eigenvalues}, which we have no knowledge except for the
ground energy $\lambda^B_0=0$.  In the second step, we relate the first
singularity of $F(k)$ to the proliferation of periodic orbits.  In fact, we
have
$$
\eqalign{
	F(k)&\sim\sum_pl(p)e^{-kl(p)}~~~~l\rightarrow\infty ,\cr
	&\sim\int x e^{-kx}d\Pi(x)~~~~x\rightarrow\infty .\cr
	}\eqn\hrokusan
$$
This implies by the above arguments
$$
	\int x e^{-kx}d\Pi(x)\sim{1\over k-\lambda_{max}}~.
	\eqn\hrokuyon
$$
If we now perform the inverse Laplace transformation, we get
$$
	\Pi(x)\sim e^{\lambda_{max}x}/x~,  \eqn\hrokugo
$$
and hence
$$
	\alpha=\lambda_{max}~.  \eqn\hrokuroku
$$
Here we stress the following points:

\item{(i)}  the asymptotic for of $\Pi(x)$ has been determined from
the real eigenvalue $\lambda_{max}$ which is quite different in comparison
to the dynamical system on the ordinary Riemann surface.  In the latter
model, the ground energy ($\lambda_0=0$) controls the exponential
proliferation of the periodic orbits through the trace formula and leads to
the asymptotic formula
$$
	\Pi(x)\sim e^x/x. \eqn\hrokunana
$$
\item{(ii)}  Our derivation includes the delicate arguments so the
rigorous proof should be developed.  One way to the direction is to
investigate the analytic properties of $L$-function defined by
$$
\eqalign{	 L(s)&=Z(s+1)/Z(s)~,\cr
	{}&Z(s)=\proprim \prod_{n=0}^{\infty}\,
	 \left(1-e^{-(s+n)l(p)}\right) ~,\cr
	} \eqn\hrokuhachi
$$
Then the following theorem is known\findref\sunada,

\noindent {\it Theorem}

If $L(s)$ satisfies the conditions:

\item{(I)} $L(s)$ converges absolutely for ${\rm Re}~(s)>\alpha,$
\item{(I$\!$I)} $L(s)$ has a meromorphic continuation onto some region
including  ${\rm Re}~(s)>\alpha,$
\item{(I$\!$I$\!$I)}  $L(s)$ has no zero point on ${\rm Re}~(s)>\alpha,$
\item{(I$\!$V)}  $L(s)$ is holomorphic on ${\rm Re}~(s)>\alpha$ and has
a simple pole at $s=\alpha$, then the asymptotic formula exists
$$
\pi(x)\sim e^{\alpha x}/\alpha x~.\eqn\hrokyu
$$

Our previous arguments support the conditions (I)-(IV), however, the
constant $\alpha$ has not been determined strictly due to
{\it small eigenvalue} problems.

	\chapter{Moduli}

	The energy spectrum is controlled by the length spectrum, or
equivalently, the periodic orbits on SRS through the trace formula and
the length of a periodic orbit is determined by the norm function of the
corresponding \sgamma-element.  This permits us to think of the {\em
moduli} for SRS which is the free parameters in \sgamma.  In the theory
of Riemann surface, it is known that some lengths corresponding to $6g-6$
primitive periodic orbits can be chosen as the moduli parameters.
The similar situation may be expected in the case of SRS.
In  Sec. 2, we saw that the moduli of SRS is the $6g-6$ Grassmann even
and $4g-4$ odd parameters and hence lengths of $6g-6$ should be used
as the even moduli parameter.  Unfortunately, we have no quantity for
the odd-moduli, i.e., the mechanical observable is Grassmann even.

We first note that the length $l(k),~k\in {\rm S}\Gamma$ is written as
$$
	l(k)=\log(k(S),S,U_k,V_k) \eqn\hnanajuu
$$
Here $U_k(V_k)$ is the repelling (attractive) fixed point of $k$
(see \juu\ and \ichiichi) and $S$ is an arbitrary point in
${\rm I\!\!R}_s-\{U_k,V_k\}$. The Grassmann even bracket of 4-points
$(Z_1,Z_2,Z_3,Z_4)$ on ${\rm SH}\bigcup {\rm I\!\!R}_s$ is defined by
$$
	(Z_1,Z_2,Z_3,Z_4)={{z_{13}z_{24}}\over{z_{14}z_{23}}},
	\eqn\hnanaichi
$$
where $Z_i=(z_i,\theta_i)$ and $z_{ij}=z_i-z_j-\theta_i\theta_j$.  This
is invariant under \SPL~ and its body is actually the ordinary cross
ratio invariant under the M{\"o}bius transformations.  Next we introduce
the Grassmann odd \SPL~invariant quantity.  This is defined by the
bracket of 3-points on SH$\bigcup{\rm I}\!\!{\rm R}_s$,
$$
(Z_1,Z_2,Z_3)={\theta_{123}\over{(z_{12}z_{23}z_{31})^{1/2}}}~~,
	\eqn\hnanani
$$
where
$\theta_{123}=\theta_1z_{23}+\theta_2z_{31}+\theta_3z_{12}+
\theta_1\theta_2\theta_3$.
The ordering of 3-points in the above formula is fixed (up to cyclic
permulations) by demanding that $(z_1-z_2)(z_2-z_3)(z_3-z_1)>0$.
Using this invariant, we can provide $4g-4$ odd moduli parameters.
Let $\{A_i,B_i\}~~(i=1\sim g)$ be the generators of \sgamma~ in
Eq.\nana, and $\{U^A_i,V^A_i\}$ and  $\{U^B_i,V^B_i\}$ be the fixed
points of the generators $A_i$ and $B_i$, respectively.  Then odd
moduli parameters $\{\lambda(k)\}~(k=1\sim 4g-4)$ are given by
\findref\oursixth,
$$
\eqalign{
	\{\lambda(k)\}=&\{(U^A_1,V^A_1,U^A_j),(U^A_1,V^A_1,V^A_j),\cr
	&(U^A_1,V^A_1,U^A_j),(V^A_1,V^A_1,V^A_j),~~j=2,3,
	\cdots h\}.\cr}\eqn\hnanasan
$$
Since the condition \nana\ on the generators is invariant under
conjugation, one may regard $A_1$ as
diagonal, i.e., the fixed points of $A_1$ can be put to
$\{(0,0),(\infty,0)\}$.  Then the condition reveals that the
parameters in $B_1$ is written by the parameters in
$\{A_j,B_j\}\ (j=2,\cdots g)$.  The $\{\lambda(k)\}$ represent
essentially all the odd parameters in the remaining generators,
$\{A_j,B_j\}\ (j=2,\cdots g)$.
Thus $\{l(i),\lambda(j)\}\ (i=1,\cdots,6g-6,~j=1,\cdots,4g-4)$ provide
the moduli of SRS.  These moduli parameters have the manifest \SPL\
invariance by the construction and offer the good coordinates on the
moduli space of SRS.

\refout
\bye